\documentclass[conference]{IEEEtran}

\usepackage{subfigure}
\usepackage[dvipsnames]{xcolor}
\usepackage{amsmath}
\usepackage{amssymb}
\usepackage{amsthm}
\usepackage{comment}
\usepackage{color}
\usepackage{mathtools}
\usepackage{algorithm}
\usepackage{algorithmicx}
\usepackage[noend]{algpseudocode}

\makeatletter
\def\BState{\State\hskip-\ALG@thistlm}
\makeatother

\usepackage{datetime}
\usepackage{graphicx}
\usepackage{cuted}
\usepackage{pgfplots}
\usepackage{filecontents}  
\pgfplotsset{compat=1.8}

\ifCLASSINFOpdf
\else
\fi
\begin{document}
%

\title{Sense-Bandits:  AI-based Adaptation of Sensing Thresholds for Heterogeneous-technology Coexistence Over Unlicensed Bands}

\author{\IEEEauthorblockN{Mohammed Hirzallah and  Marwan Krunz 
}

\IEEEauthorblockA{$^1$Department of Electrical and Computer Engineering, University of Arizona, AZ, USA\\
Email: \{hirzallah, krunz\}@email.arizona.edu} 
}

%


\maketitle


\begin{abstract}
In this paper, we present \emph{Sense-Bandits}, an  AI-based framework for
distributed adaptation of the sensing thresholds (STs) over shared spectrum. This framework specifically targets the coexistence of 
heterogenous technologies, e.g., Wi-Fi, 4G Licensed-Assisted Access (LAA), and 5G New Radio Unlicensed (NR-U), over unlicensed channels.
To access the channel, a device compares the measured power  with a predefined ST value and accordingly decides if the channel is idle or not. Improper setting of the 
ST values creates asymmetric sensing floors, resulting in collisions due to  hidden terminals  and/or reduction in the spatial reuse due to  exposed  terminals. Optimal ST setting is challenging because it requires global knowledge of mobility, traffic loads, and channel access behavior of all contending devices. \emph{Sense-Bandits} tackles this problem by employing a 
clustering-based multi-armed bandit (MAB) algorithm, which  adapts its learning behavior based on network dynamics.  Clustering allows the algorithm to track network changes in real-time, ensuring 
fast learning of the best ST values by classifying the state and dynamics of coexisting networks. We  develop a C++-based network simulator that implements  \emph{Sense-Bandits} and we apply
it to evaluate the coexistence of Wi-Fi and 5G NR-U systems over the unlicensed 5 GHz U-NII bands. Our simulation results indicate that ST-adaptive devices employing \emph{Sense-Bandits} do not harm neighboring devices that  
 adopt a fixed ST value. 
\end{abstract}


%

\section{Introduction}
5G and beyond  cellular networks will host new applications with very demanding requirements in terms of extremely high throughput  
as well as ultra-reliable and low-latency transmissions \cite{ITU-IMT2020-Chairman-Notes}. Licensed spectrum below 7 GHz is overly subscribed and cannot meet the demands of these new applications. 
Additional spectrum is needed. Although the recently opened  millimeter-wave (mmWave) bands offer an abundance of spectrum, transmissions over 
these bands are highly susceptible to blockage and require complex beamforming/tracking procedures to maintain directional line-of-sight communications.
Unlicensed bands below $7$ GHz, on the other hand, offer about $2$ GHz of spectrum. They are partially used by IEEE 802.11-based systems (e.g., Wi-Fi). Mobile network operators (MNOs) are pushing for making these bands available for cellular operation under  4G Licensed Assisted Access (LAA) and, more recently, 
5G New Radio Unlicensed (NR-U) \cite{TR36.889LAA}\cite{TR38.889-nru-2}. Both technologies would use the unlicensed sub-7 GHz spectrum to supplement their licensed services. NR-U can also be leveraged in a standalone mode, offering great opportunity to new MNOs who do not have their own licensed spectrum \cite{Lagen2018nru-mmwave, Hirzallah2020TCCN-nru}.

Unlicensed-band cellular operation, including LTE-LAA and 5G NR-U, comes with its own  challenges. Chief among them is fair coexistence with other incumbents,  particularly Wi-Fi systems. 
Fair access to the shared channel is strongly dependent on the  setting of the sensing threshold (ST), which also affects  spatial reuse and  channel utilization. 
To access an unlicensed channel, a device, e.g., Wi-Fi access point (AP), Wi-Fi station (STA), 5G NR-U user equipment (UE), etc., must first execute a Listen-Before-Talk (LBT) procedure, which is 
essentially a variant of the CSMA/CA protocol.  According to LBT, the device senses the channel and compares the received power with a predetermined ST value to determine if the channel is idle. NR-U/LAA and IEEE 802.11-based technologies adopt different ST settings that result in heterogeneity of their sensing floors. This creates  hidden and exposed nodes, which  reduce 
the total network throughput \cite{Jiang2007tmc-hidden-exposed-node}. For example, IEEE 802.11n/ac-based Wi-Fi devices are set with an energy-based ST value of $-62$ dBm to detect unknown signals and a signal-based ST value of $-82$ dBm to detect Wi-Fi signals. The IEEE 802.11ax-based devices are set with an additional signal-based ST value, which is variable, to detect Wi-Fi signals coming from overlapping Basic Service Sets (OBSS). LAA and NR-U, on the other hand, adopt a fixed maximum energy-based ST value of $-72$ dBm. Optimal setting of ST values requires real-time global knowledge about the wireless environment as well as dynamics of neighboring networks, including their mobility and traffic loads. 

To maximize network performance, we propose a distributed framework that allows every device to learn its optimal ST value in an online manner with the least possible communication overhead. Employing AI and machine learning (ML) can significantly harmonize coexistence between homogeneous and heterogeneous wireless systems \cite{Hirzallah2017jsac,Amir2019Dyspan,Hirzallah2019Secon-matchmaker,Hirzallah2020icnc}. Our framework, called \emph{Sense-Bandits}, is based on a multi-armed bandit (MAB) algorithm. MAB algorithms are  a class of 
reinforcement learning (RL) that aim at  establishing a balance between exploitation (i.e., maintaining the current ST value) and exploration (i.e., trying new ST values) by 
minimizing the accumulated regret between actual and optimal rewards. However, traditional MAB algorithms suffer from a long convergence time and  inability 
to cope with fast-varying environments, a.k.a., \emph{cold start}. For example, when the environment experiences fast changes, the learning outcome 
of the MAB algorithm becomes outdated and misleading to the adaptation process. To cope with this challenge,  \emph{Sense-Bandits}
allows devices to detect changes in the environment dynamics and \emph{re-initiate} the learning process properly by setting the initial values of different arms (actions). 
This re-initialization utilizes the history of previous experience and copes with the rate at which the environment changes.
To make the adaptation process aware of real-time changes, each device constructs over time a feature representation called  of the environment a \emph{sensing fingerprint (SF)}. The SFs  
are used to track the environment and detect real-time changes by finding the similarity between SFs collected over consecutive time periods. Once a change in the environment is detected, the device follows a rule to decide whether the most recent learning outcomes are outdated and whether the device needs to re-initiate its learning process. To reduce the overhead of re-initializing the learning process, we let the device exploit its previous/stored history of learning experience. This is achieved by finding states that are already learned and have close similarity with the new state of the environment. 
Sense-Bandits also considers the important tradeoff between utilizing historical learning experiences and limiting the  storage and computational requirements at a device. Specifically,
the environment may give rise to a huge number of states, making the learning experience of these states computationally prohibitive. By completely ignoring the learning experience, we run into 
the shortcomings of traditional MAB algorithms. On the other hand, maintaining the complete history of the learning experience incurs high memory and computational overheads.
In Sense-Bandits, we focus on finding a balance between these two extremes  by clustering the history of learning instances into a finite number of clusters based on the similarity of their underlying states. Our clustering-based multi-armed bandit (CMAB) algorithm accommodates this balance by finding the nearest cluster that has a similar state to the new environment. We associate with every cluster a recommended initial values of STs. 
Compared with traditional MAB algorithms, CMAB has a faster convergence and can cope better with various network dynamics. 

The rest of the paper is organized as follows. In Section \ref{sc:related-work}, we review  related works on ST adaptation. The system model and problem formulation are discussed in
Section~\ref{sc:sensecontrol-framework}. Sense-Bandits is presented in Section~\ref{sec:senseBandits}. Evaluation of Sense-Bandits in a simulated Wi-Fi / 5G NR-U coexistence scenario
is presented in Section~\ref{sec:evaluation}, followed by conclusions in Section~\ref{sc:conclusions}. Throughout the 
paper, we use the superscript $(\cdot)^{(u)}$ to refer to NR-U and $(\cdot)^{(w)}$ to refer to Wi-Fi.

\section{Related Work} \label{sc:related-work}
 Adapting STs for  harmonious coexistence between LAA and Wi-Fi systems has been previously investigated. 
 Li et al. \cite{Li2017tvt-laa-ed-adaptation} investigated adapting the ST value for LAA devices by increasing or reducing it based on the collision rate. 
 Iqbal et al. \cite{Iqbal2017laa-ed-ns3} investigated the impact of changing the ST values in LAA and Wi-Fi devices, and concluded that lowering these values for Wi-Fi could improve the throughput of both networks. 
 Ajami et al. \cite{Ajami2017twc-11ax-stochastic-geometry} analyzed the coexistence between LTE and Wi-Fi systems using stochastic geometry, and suggested that LTE could be a good neighbor to Wi-Fi if the later adapts its ST. Mehrnoush et al. \cite{Mehrnoush2018tnet-laa-modeling} modeled the impact of changing the ST values using Markov-models. While these  works offer great insights into ST adaptation,
 the approaches are either ad hoc or  focus mainly on evaluating the effect of changing the ST value rather than adapting it online.
 
 Adapting ST values to improve the spatial frequency reuse has also been discussed in the context of  IEEE 802.11 networks. Most of the works assume homogeneous devices (e.g., Wi-Fi) and require
 these devices to be able to decode   certain fields, e.g., Basic Service Set (BSS) color bit. Kulkarni et al. \cite{kulkarni2016dynamic} presented  extensive evaluation to show the impact of adapting 
 the ST values on improving IEEE 802.11ac network throughput. In \cite{Mvulla2018-access-dual-sensing}, the authors investigated improving IEEE 802.11ax network performance 
 by using dual ST values, one is conservative and targets the detection of intra-BSS signals while the other  is aggressive and targets the detection of
inter-BSS signals. The authors in \cite{Oteri2015globecom-spatial-reuse} demonstrated that adapting the ST values is needed and the best ST value is scenario-dependent. 
Afaqui et al. \cite{afaqui2016globecom-cca-adaptation} proposed a framework to adapt the ST values in IEEE 802.11ax networks based on the received interference generated by nearby APs. 
Selinis et al. \cite{selinis2019damysus} presented a framework, called Damysus, for adapting ST values based on the color bit. Recently, researchers started investigating the use of learning techniques 
to adapt the ST value in IEEE 802.11ax networks.  The authors in \cite{wilhelmi2019-ml-cca} presented a MAB-based framework to the control ST value, transmit power, and channel selection so as to  improve
the  spatial reuse over IEEE 802.11ax networks. They also demonstrated the advantage of MAB in achieving collaboration among APs 
to efficiently adapt their ST values \cite{wilhelmi2019collaborative}. Although these works present exciting results, they are still focused on a homogeneous technology whereby devices can 
decode each other's frames and read, for example, the color bit to distinguish between signals coming from different BSSs. These assumptions are not applicable to heterogeneous coexistence setting, e.g., NR-U and Wi-Fi coexistence. In contrast, our Sense-Bandits approach
  is technology-agnostic and can be applied to all technologies sharing an unlicensed channel.

\section{System Model and Problem Formulation}  
\label{sc:sensecontrol-framework}
Without loss of generality, we consider two coexisting networks: A 
5G  NR-U network that  consists of a set $\mathcal{B}  = \{B_1, \cdots, B_{N_b}\}$ of $N_b$ base stations (BSs) and serves a set $\mathcal{U} = \{U_1, \cdots, U_{N_u}\}$ of $N_u$ UEs,
and a  Wi-Fi network that  consists of a set $\mathcal{P} =\{P_1, \cdots, P_{N_p}\}$ of $N_p$ APs and serves a set $\mathcal{S} = \{S_1, \cdots, S_{N_s}\}$ of $N_s$ STAs. 
Let $\mathcal{N}  = \mathcal{U} \cup \mathcal{B}$  and let $\mathcal{W} = \mathcal{S} \cup \mathcal{P}$. UEs (STAs) attach to the BS (AP) that provides the strongest signal. 
Our formulations and evaluations are based on single-antenna devices, but can be easily extended for MIMO operation. 
5G NR-U and Wi-Fi network share an arbitrary unlicensed channel of bandwidth $W_c$ in Hz.  An LBT procedure (CSMA/CA with exponential backoff) is used for channel access at a timing 
granularity of $\Delta_c$, where $\Delta_c = 9$ microseconds corresponds to the duration of a MAC time slot. This is inline with NR-U specifications and IEEE 802.11 standards for operating over
 the 5 GHz UNII bands \cite{TS37.213nru}\cite{IEEE802.11-2016}. 

To ensure the channel is idle, a device compares the sensed signal power with a predefined ST value, a.k.a., detection threshold or clear channel assessment (CCA) threshold. If the channel remains idle for a period of time, a.k.a., Arbitration Inter-Frame Space (AIFS) (or Initial Deferment period), the device starts transmission; otherwise, it initiates a counter with a random value $k$ and backs off 
for $k$ idle time slots, where:
  \begin{align} \label{eq:backoff-duration}
  k \sim \mbox{uniform}\{0,\cdots, \min (2^\varsigma W_{\min}, W_{\max})-1\}
  \end{align}
$W_{\min}$ and $W_{\max}$ are the minimum (CWmin) and maximum (CWmax) size of the contention window, respectively, and $\varsigma$ is the retransmission attempt. Let $W^{(u)}_{\min}$, $W^{(u)}_{\max}$, and $a^{(u)}$ be the CWmin, CWmax, and AIFS values adopted in the NR-U network, respectively. $W^{(w)}_{\min}$, $W^{(w)}_{\max}$, and $a^{(w)}$ are defined similarly for the Wi-Fi network. 
During the backoff time, if the channel becomes busy, the device must freeze its counter and await for the channel to become idle again. The device can only decrease its counter if the channel is deemed to be idle.
This decision clearly depends on the ST value. Let $\gamma^{(u)}_j$ be the ST value adopted by NR-U device $j \in \mathcal{N} $ and $\gamma^{(w)}_l$ be the ST value adopted by Wi-Fi device $l \in \mathcal{W}$. In our work, we consider a range of  $\gamma^{(u)}_j$ and $\gamma^{(w)}_l$ values that covers the ST values adopted in the standards, including both energy and signal detection thresholds. 
The sensed signal power $y^{(u)}_i(n)$ at an arbitrary time slot $n$ by an arbitrary UE $U_i$ is given by (similar expressions can be formulated for APs, STAs and BSs):  
\begin{align}
y^{(u)}_i (n) &= \!\!\!\sum_{\underset{ j \neq i}{j \in \mathcal{N}}}\! \Upsilon_j^{(u)} (n) h_{ji} (n)|s^{(u)}_j (n)|^2 \mathbf{1}^{(u)}_j (n) \nonumber \\
 &+\!\!\! \sum_{l \in \mathcal{W}}\!\! \Upsilon^{(w)}_l(n) h_{li}(n) |s^{(w)}_l(n)|^2 \mathbf{1}^{(w)}_l (n)  + z^{(u)}_i(n) \label{eq:rrs-nru}
\end{align}
where $\Upsilon_l^{(w)}(n)$ and $\Upsilon_j^{(u)}(n)$ are the transmit power of an arbitrary Wi-Fi device $l$ and NR-U device $j$ at time $n$, $s_j^{(w)}(n)$ and $s_l^{(u)}(n)$ are the transmit signals of 
an arbitrary Wi-Fi device $l$ and NR-U device $j$ at time $n$, $h_{ji}(n)$ is the channel gain between transmitting device $j$ and receiving device $i$ at time $n$, respectively, and $z^{(u)}_i(n)$ is the additive 
white Guassian noise at NR-U device $i$. $\mathbf{1}^{(u)}_j (n)$ and $\mathbf{1}^{(w)}_l (n)$ are the indicator functions that NR-U device $j $ and Wi-Fi device $l$ access the unlicensed channel at an arbitrary time $n$, respectively:
\begin{align}
\mathbf{1}^{(u)}_j (n) = \{1: \mbox{if } y^{(u)}_j(n) \leq \gamma^{(u)}_j \text{ and } k^{(u)}_j(n) = 0 \} \\
\mathbf{1}^{(w)}_l (n) = \{1:  \mbox{if } y^{(w)}_l(n) \leq \gamma^{(w)}_l \text{ and } k^{(w)}_l (n) = 0 \}
\end{align}
where $k^{(u)}_j(n)$ and $k^{(w)}_l(n)$ are the backoff counter for NR-U device $i$ and Wi-Fi device $l$ at time slot $n$, and they are initialized as in (\ref{eq:backoff-duration}). There are several factors that affect the setting of these indicator functions, including the ST value used by neighboring devices as well as their channel access behavior, traffic loads, and mobility patterns \cite{Hirzallah2019tccn}. Expressing these indicator functions based on these factors requires notoriously complicated stochastic-geometry analysis \cite{Ajami2017twc-11ax-stochastic-geometry}.

\subsection{Problem Formulation} \label{sc:optimum-sensitivity-threshold}
At time $n$,  the  uplink throughput $S^{(u)}_j(n)$ for an arbitrary UE $U_j$ attached to an arbitrary BS $B_i$ can be expressed as (similar expressions can be formulated for NR-U downlink as well as Wi-Fi uplink and downlink communications):
\begin{align} \label{eq:sinr}
S^{(u)}_j(n) = W_c\: \mathbb{E}\Big[\log (1+ \frac{Y_{i} (n)}
{ I_{i}(n) + z^{(u)}_i(n)} )\Big]
\end{align}
where  $Y_{i} (n)$ and $I_{i} (n)$ are the received signal power and interference power received by BS $B_i$, respectively:
\begin{align}
Y_{i} (n) &= \Upsilon^{(u)}_j(n) \: h_{ji} (n) \:|s^{(u)}_{j} (n)|^2 \: \mathbf{1}^{(u)}_j(n) \\
I_{i} (n) &= \sum_{\ell\in \mathcal{N}, \ell \neq j} \Upsilon^{(u)}_\ell (n) \: h_{\ell i}(n) \: |s^{(u)}_{\ell}(n)|^2 \: \mathbf{1}^{(u)}_\ell (n) \nonumber \\
&+ 
\sum_{l \in \mathcal{W}} \Upsilon^{(w)}_l (n)\: h_{li}(n)\:|s^{(w)}_{l}(n)|^2\: \mathbf{1}^{(w)}_l (n).
\end{align}
 The expectation in  (\ref{eq:sinr}) accounts for the randomness in the interference generated by neighboring devices due to their mobility, traffic loads, and channel access behavior. Our objective is to maximize the sum-throughput experienced by both NR-U and Wi-Fi devices over a period of $T_n$ time slots:
\begin{align}
P_1: \underset{\Gamma^{(u)}(n),\Gamma^{(w)}(n) }{\arg \max} & \:\: \sum_{n=1}^{T_n} \Big[\sum_{j \in \mathcal{N}} S^{(u)}_j(n)  + \sum_{l \in \mathcal{W}} S^{(w)}_l(n)\Big] \label{eq:objective-function} \\
 \text{s.t.}\:\: & \:\:  \gamma^{(u)}_j (n), \gamma^{(w)}_l (n) \in \Gamma,\:\: l \in \mathcal{W}, j \in \mathcal{N} \label{eq:constraint-wifi}
\end{align}
where $S^{(w)}_l(n)$ is the throughput achieved by Wi-Fi device $l$ at time $n$,  $\Gamma = \{\gamma_1, \cdots, \gamma_{N_a}\}$ is the set of possible ST values that can be selected by NR-U and Wi-Fi devices, $\Gamma^{(u)}(n)= \{\gamma^{(u)}_j(n) | j \in \mathcal{N}\}$, and $\Gamma^{(w)}(n)= \{\gamma^{(w)}_l(n) | l \in \mathcal{W}\}$. The decision variables are the ST values $\Gamma^{(n)}(n)$ for NR-U 
devices and $\Gamma^{(w)}(n)$ for Wi-Fi devices. The above optimization is stochastic and nonlinear. In principle, it can be solved via dynamic programming. However, such an approach
gives rise to several challenges: 
\begin{itemize}
\item Expressing the objective function as a function of ST values is mathematically intractable and putting it in a closed form is not possible.

\item Solving the dynamic programming problem  requires global knowledge about the  dynamics of NR-U and Wi-Fi networks, including channel conditions, location information, traffic loads, 
the state of  backoff counters, etc. Obtaining this global knowledge is practically difficult due to the large communication overhead and privacy concerns.

\item Problem $P_1$ involves taking decisions over time, and solving this problem in a distributed fashion requires coordination and synchronization between devices. Due to the differences in the 
waveforms and message formats between Wi-Fi and 5G NR-U, achieving this coordination/synchronization is difficult.
\end{itemize}
 In contrast to  an exact dynamic programming approach, reinforcement learning offers heuristics, including MABs, to solving $P_1$ by employing a learning agent. 
 Specifically, in a MAB, the learning agent aims at finding a balance between exploiting ST values of known average reward and exploring new ones. The learning process focuses on minimizing the accumulated regret expressed by the difference between the average rewards (e.g., throughput)  of the optimal and actual ST values. Most well-known MAB algorithms, e.g., upper confidence interval (UCB), epsilon greedy, Thompson sampling, etc., are subject to long learning and convergence times. In a time-varying environment, such as a dynamic unlicensed channel, these algorithms  fall short of approaching the optimal solution. 
 Sense-Bandits overcomes this limitation  and ensures fast adaptation by employing clustering of the environment and running a cluster-based MAB (CMAB) algorithm. This 
 CMAB algorithm gives the learning agent awareness of the channel and network dynamics, and provides it with prior information required to speed up the learning process.  

\section{Sense-Bandits Design}
\label{sec:senseBandits}
In Sense-Bandits, an LBT device (e.g., NR-U or Wi-Fi device) runs its own learning gent and takes actions in a distributed fashion. For ease of illustration, we drop the subscript that denotes the device index and explain the formulation from a single device perspective. Our formulation applies to NR-U and Wi-Fi devices. In our subsequent notation, we let the first subscript index denote the time and 
the second subscript index denote the state of the environment or action, as  applicable. 

\subsection{Learning Model}
\subsubsection{Time Horizon}

We consider a finite-time horizon $\mathcal{T} =  \{1,\cdots, T\}$ that consists of $T$ time epoch. We also use $t \in \mathcal{T}$ to denote the index of time epochs. Every epoch has a fixed duration of $\Delta_p$ seconds. Selection of ST value should take place at the start of the time epoch, and during this time epoch the device collects observations and monitors the achieved performance. These observations and performance are used to trigger the selection of ST value to be used over the next time epoch.

\subsubsection{State of Environment}
The state of the environment represents the unlicensed wireless environment, including dynamics of coexisting networks, such as their location information, mobility pattern, traffic loads, channel access behavior, etc.  Wi-Fi and NR-U devices should select their best ST values independently based on how they view the state of the wireless environment with the least communication overhead possible. Achieving global knowledge about the state of environment by NR-U and Wi-Fi devices is difficult due to the challenges discussed before. Therefore, instead of obtaining a global knowledge, every device constructs a  \emph{sensing fingerprint} (SF) in which a normalized histogram of the sensed signal powers is constructed over a monitoring period $\Delta_m$, where $\Delta_m \leq \Delta_p$. For example, the monitoring period of the $t$th epoch can be started in the middle of $t$th time epoch. In Figure \ref{fig:cmab-updates}, we show an example in which the start of monitoring period aligns with epoch boundary. The monitoring period is divided to sensing periods $\Delta_s$ in which the sensed signal power is to be computed, where $\Delta_s \ll \Delta_m$. One option is to set the sensing period to be equal to MAC time slot $\Delta_s$, i.e., $\Delta_s = \Delta_c = 9$ microseconds. The sensing is supposed to take place during the backoff process and to be suspended during the transmission time. Let $e_{t} =\langle \tilde{e}_{t,1}, \cdots, \tilde{e}_{t,N_r} \rangle$ be the normalized histogram that represents the state of the environment as observed by the device at time epoch $t$, where $N_r$ is the number of bins and $\tilde{e}_{t,j} \in [0,1]$ is the probability that the sensed signal power is in the $[ \delta_j, \delta_{j+1})$ interval. Note that $e_{t}$ is a simplex in $\mathbb{R}^{N_r}$, i.e., $\sum_{j=1}^{N_r}\tilde{e}_{t,j} = 1$. The overhead required to construct the state of the environment is practical because both 3GPP specifications and IEEE 802.11 standards require wireless devices to track statistics of their sensed signal powers and report them back to base station and access point. In our work, devices will utilize their sensed signal power locally and they are not supposed to exchange it with their home base station or access point. Characterizing the state of the environment using the SF profile takes advantage of the sensing part that happens during the backoff process, and thus it requires no significant overhead.

\subsubsection{Actions}
We let $\Gamma = \{\gamma_1, \cdots, \gamma_{N_a}\}$ be the set of $N_a$ possible actions that represent the ST values that a device can use while contending for a channel access. Let $a_{t}$ be the ST value, i.e., action or arm, that the device uses while contending for a channel access during time epoch $t$, where $a_{t} \in \Gamma$. Let $x_{t,a}$ be the action vector at time epoch $t$, where $||x_{t,a}||_2 \leq d_x$, $d_x = 1$. In our work, actions are selected with hard decision. The size of $x_{t,a}$ is equal to the number of actions, i.e., ST values to be considered as actions. For example, when $\gamma_j$ is selected, then the $j$th element in $x_{t,a}$ will be set to one, while other elements are set to zeros. 
\subsubsection{Rewards}
We let the effective throughput achieved during a time epoch be the reward observed by the learning agent. The effective throughput per epoch can be calculated by taking the difference between successful and failed traffic exchanged in one epoch divided by its duration.  It is clear that the reward received at a particular time epoch is random and depends on the state of the environment, including ST values, i.e., actions, taken by different devices sharing the same channel. Let $R_{t,e,a}$ be a random variable that represents the reward received by the device, expressed by the effective throughput achieved during time epoch $t$, for taking an action $a$ while the state of the environment is $e$ \footnote{For latency stringent applications, the reward can be formulated based on latency experienced during time epoch $t$, and thus the problem in  (\ref{eq:objective-function}) becomes a maximization of the summation of the inverse of delays.}. For ease of illustration, we drop the time index in the subscript of $e$ and $a$, and time information can be inferred from their associated quantity. Let $r^{(s)}_{t,e,a}$ and $r^{(f)}_{t,e,a}$ be the amount of traffic (expressed in bits) exchanged with success, i.e., with ACK, and with failure, i.e., NACK/ACK-timout, respectively, during time epoch $t$  while taking an action $a$. The sampled reward, $r_{t,e,a} \sim R_{t,e,a}$, as observed by the learning agent can be expressed as:
\begin{align} \label{eq:observed-reward}
r_{t,e,a} \leftarrow (r^{(s)}_{t,e,a} - r^{(f)}_{t,e,a})/\Delta_p
\end{align}

It is a common practice to model the random reward, $R_{t,e,a}$, by a linear function, a.k.a., \emph{linear bandits} \cite{abbasi2011improved-MAB-bounds}. The reward at time epoch $t$ can be modeled using the following linear relation:
\begin{align} \label{eq:reward}
R_{t, e, a}  \doteq \mu_{t,e}^\top   x_{t,a} + \epsilon_{t,e}
\end{align}   
where $\mu_{t,e}$ is a vector that represent \emph{actions utility} for state $e$, i.e., the average rewards of actions when the environment is at state $e$, where $||\mu_{t,e}||_2 \leq d_{\mu}$. In other words, the $i$th element in $\mu_{t,e}$ represents the average reward for the $i$th action. Actions utility $\mu_{t,e}$ is unknown to the learning agent because of the noise $\epsilon_{t,e}$, where noise has a zero mean and $\nu$-sub-Gaussian tail: 
\begin{align} \label{eq:sub-gaussian-noise-parameters}
\forall \lambda \in \mathbb{R},\qquad \mathbb{E}_{\epsilon_{t,e}}[e^{\lambda \epsilon_{t,e}} | \textbf{x}_{1:t}, \textbf{e}_{1:t}] \leq \exp(\lambda^2\nu^2/2)
\end{align}
where $\lambda$ and $\nu$ are the parameters of the $\nu$-Sub Gaussian noise $\epsilon_{t,e}$, $\mathbf{x}_{1:t}$ is the sequence of previous action history and $\mathbf{e}_{1:t}$ is the sequence of previous noise history. It is obvious that the mean of reward is $\mu_{t,e}^\top   x_{t,a}$.

\subsection{Selection of Optimal ST Value}
Let $a^{*}_t$ be the optimal ST value that the device should select at time epoch $t$ to maximize its expected reward at time $t$. The optimal action $a^{*}_t$ can be expressed as follows:
\begin{align} \label{eq:optimal-max-action}
a^{*}_t  = \underset{a \in \Gamma}{\arg\max} \:\: \mathbb{E}[R_{t,e,a}] = \underset{a \in \Gamma}{\arg\max} \:\: \mu_{t,e}^\top   x_{t,a}
\end{align}
The learning agent does not know the true action utility, i.e., $\mu_{t,e}$, and hence it incurs a regret due to selection of a non-optimal ST value, say $a \neq a^*$. Let $g_t$ be the expected regret at time epoch $t$ that can be expressed as:
\begin{align} \label{eq:exp_regret}
g_t = \mathbb{E}[R_{t,e,a^*} - R_{t,e,a}] = \mu_{t,e}^\top   x_{t,a^*}  - \mu_{t,e}^\top   x_{t,a}
\end{align} 
We define the \emph{accumulated regret}  $G(T)$ up until time epoch $T$ to be the sum of expected regrets:

\begin{align} \label{eq:exp_accumulated_regret}
G(T) =  \sum_{t = 1}^{T}g_t =\sum_{t=1}^{T} \mu_{t,e}^\top   x_{t,a^*}  - \mu_{t,e}^\top   x_{t,a} 
\end{align} 
where the expectation is taken over the randomness of environment states and rewards as expressed by the $H_t$-conditioned history, where $H_t = \{e_1, x_{1,a}, r_{1,e,a}, \cdots, e_{t-1}, x_{t-1,a}, r_{t,e,a}, e_t\}$ is the sequence of state-action-reward history observed and taken by the device up until time epoch $t$. 

The problem of finding the optimal actions in (\ref{eq:optimal-max-action}) is equivalent to finding the sequence of actions, i.e., $\{a_1, a_2, \cdots, a_T\}$ that minimizes the accumulated expected  regret over the $T$ epochs:
\begin{align} \label{eq:optimal-action}
\underset{\{a_1, \cdots, a_T\}}{\arg\min} \:\:G(T) \\
s.t.,& \:\: a \in \Gamma \nonumber
\end{align}

\subsection{Clustering of States}

We consider a time variant wireless environment where the state of the environment follows a time variant distribution $f_{t,e}$, i.e., $e_{t} \sim f_{t,e}$. We assume the learning agent is unaware of this distribution. We let $E = \{e_t| e_t \in \mathbb{R}^{N_r}, \sum_{j=1}^{N_r}\tilde{e}_{t,j} = 1\}$ be state space of the environment. We partition the states of environment in $E$ to $N_c$ subsets, i.e., clusters, based on their similarities. Let $\mathcal{C} = \{E_1, \cdots, E_{N_c}\}$ be the  set of the $N_c$ disjoint clusters, where $E=\cup_{k=1}^{N_c}E_k$ and $E_{k} \cap E_{l} = \emptyset$. For each cluster, say $E_k$, we define a centroid, $\bar{e}_k$, radius, $\bar{d}_k$, and average actions utility, $\bar{\mu}_k$. One possible approach to cluster states is using K-mean algorithm where the distance between states is expressed based on a similarity measure such as Kullback-Leibler (KL) divergence. Let $e_{t}= \{\tilde{e}_{t,1}, \cdots, \tilde{e}_{t,N_r}\}$ and $e_{t^\prime} = \{\tilde{e}_{t^\prime,1}, \cdots, \tilde{e}_{t^\prime,N_r}\}$ be two arbitrary states captured over two different time epochs, i.e., $t$ and $t^\prime$. The KL divergence between these two states can be expressed as:
\begin{align}
V_{KL} (e_{t} || e_{t^\prime}) \doteq  \sum_{l= 1}^{N_r} \tilde{e}_{t,l} \log (\tilde{e}_{t,l}/\tilde{e}_{t^\prime,l})
\end{align}

We can also define \emph{average actions utility} of a cluster based on the utilities of its constituting states. Let $M_c = \{\bar{\mu}_1, \cdots, \bar{\mu}_{N_c}\}$ be the set of actions utility of $N_c$ clusters, where the average actions utility $\bar{\mu}_k$ of cluster $E_k$ can be expressed as the average sum of $\mu_e$'s of states in $E_k$:
\begin{align}
\bar{\mu}_k  = \sum_{e \in E_k} \mu_e/|E_k|
\end{align}

Each cluster can be expressed by its centroid, i.e., center, and its maximum radius. Let $C = \{\bar{e}_1, \cdots, \bar{e}_{N_k}\}$ be the set of centroids of the $N_c$ clusters, where the centroid $\bar{e}_k$ of cluster $E_k$ can be expressed as the average sum of states in $E_k$:
\begin{align}
\bar{e}_k = \sum_{j}^{|E_k|} e_j/|E_k|, \qquad e_j  \in E_k
\end{align}

 Let $D = \{\bar{d}_1, \cdots, \bar{d}_{N_c}\}$ be the set of radii of $N_c$ clusters. The radius of a cluster can be expressed using KL divergence by finding the maximum divergence between states in cluster and its centroid $\bar{e}_k$:
\begin{align}
\bar{d}_k = \max \:\: \{V_{KL}(e_j||\bar{e}_k) : e_j \in E_k \}
\end{align}

Finding average actions utilities, $M_c$, set of centroids, $C$, and set of radii, $D$, of clusters can be done offline after capturing large number of measurements over diverse set of scenarios. These parameters can be loaded to mobile devices, i.e., UEs and STAs, as lookup tables. It is also possible to let base station and AP construct these parameters online and have them shared with mobile devices through signaling. It should be noted that average action utilities of different clusters, $M_c$, can be formulated based on previous learning histories acquired by running traditional MAB algorithm over large number of empirical and/or simulated environments.  
\subsubsection{Detecting Changes in Environment} 
Devices can detect substantial changes happening in the environment by tracking the best cluster that describes the environment. To track the best cluster over time, a device should continuously monitor the state $e_t$ of the environment and find the nearest cluster, $E^*_t$, as follows:
\begin{align} \label{eq:k-mean}
E^*_t = \underset{E_l \in \mathcal{C}}{\arg\min} 
\ V_{KL} (e_t||\bar{e}_l) 
\end{align}
This process has a linear complexity in terms of the number of clusters $N_c$, and it becomes expensive when the number of clusters becomes large. A better way to detect changes happening in the environment is to let the device run a detection rule in which it checks whether it deviates from its current cluster. This can be achieved by leveraging the radius of the cluster. Device can compute the similarity of its most recent state $e_t$ with the centroid of the current cluster. Let $\bar{e}_t$ and $\bar{d}_t$ be the centroid and radius of the current cluster at time epoch $t$, respectively. Let $\theta_t$ be the indicator function that signals a change in the environment at time $t$, then $\theta_t$ can be expressed as follows:
\begin{align}\label{eq:environment_change_dedetection_rule}
\theta_t = \{1: V_{KL}(e_{t-1} || \bar{e}_{t-1}) > \bar{d}_{t-1}\}
\end{align}

When $\theta_t$ is set, the device can search for the best cluster as expressed in (\ref{eq:k-mean}). This will ensure that the detection of changes happening in the environment is of low computational complexity.

\subsubsection{Reward Formulation Based on Clustering}
The linear reward formulation expressed in (\ref{eq:reward}) can be reformulated to take advantage of clustering. Let $E_k$ be the best cluster at time epoch $t$, i.e., $E^*_t = E_k$. Let $\bar{\mu}^*_{t}$ be the average action utility of cluster $E_t$ at time epoch $t$, then $\bar{\mu}^*_{t} = \bar{\mu}_k$. The random reward $R_{t,e,a}$ can be re-expressed as follows:
\begin{align} \label{eq:reward-with-clustering}
R_{t,e,a} \doteq \bar{\mu}_t^{*\top} x_{t,a} + \Delta\mu_{t,e}^\top x_{t,a} + \tilde{\epsilon}_{t,e} 
\end{align}
where $\tilde{\epsilon}_{t,e}$ is the reward noise of cluster $E_t^*$, i.e., $\tilde{\epsilon}_{t,e} = \tilde{\epsilon}_{k}$, and $\tilde{\epsilon}_{k}$ is the noise of cluster $E_k$. The noise per cluster $\tilde{\epsilon}_{t,e} $ could have different statistics from $\epsilon_{t,e} $ in  (\ref{eq:reward}), and this depends on the clustering process. Rather than learning the action utility $\mu_{t,e}$ in  (\ref{eq:reward}), the learning 
agent is supposed to learn the fractional action utility $\Delta  \mu_{t,e}$ because it already knows $\bar{\mu}^*_t$. In other words, knowledge about $\bar{\mu}^*_t$ would help in reducing the impact of noise and have more confidence about the true action utility. The noise of cluster can have similar formulation as in (\ref{eq:sub-gaussian-noise-parameters}) but with different $\lambda_k$ and $\nu_k$ parameters.

\subsection{Clustering-Multi-Armed Bandits (CMAB) Algorithm}
We next explain the design and flow of CMAB algorithm. Let $e$ be the state for which its actions utility vector needs to be learned, and let $E^*_t$ be the best cluster that represents the state of the environment. The goal is to find an estimate for $\mu_{t,e}$ based on the initial knowledge $\bar{\mu}^*_t$. In other words, the learning agent seeks to find an estimate $\tilde{\mu}_{t,e}$ for the fractional actions utility $\Delta \mu_{t,e}$ in (\ref{eq:reward-with-clustering}). This estimation can be facilitated using a procedure of online linear regression, where $\tilde{\mu}_{t,e}$ can be considered as a standard linear least-squares approximation of $\Delta \mu_{t,e}$. The estimate $\tilde{\mu}_{t,e}$ needs to be updated over time based on observations seen by the learning agent. To update the $\tilde{\mu}_{t,e}$ over time, we need to consider two entities $X_t$ and $b_{t,e}$. The $X_{t,e}$ is a matrix used to keep track of the usage of different actions over time, and it can be formulated as:

\begin{align} \label{eq:cmab-X-update}
X_{t,e} = X_{t-1,e} + x_{t-1,a} x_{t-1,a}^\top
\end{align}
where $X_{t,e}$ can be initiated to an identity matrix. The $b_{t,e}$ is a vector that tracks the accumulated rewards observed over time, and it can be updated based on the reward observed over previous epoch $r_{t-1,e,a}$ in (\ref{eq:observed-reward}) as follows:
\begin{align} \label{eq:cmab-b-update}
b_{t,e}  = b_{t-1,e} + r_{t-1,e,a}x_{t-1,a} 
\end{align}
and the initial value of $b_{t,e}$ can be set to a zero vector. Based on $X_{t,e}$ and $b_{t,e}$, $\tilde{\mu}_{t, e}$ can be expressed as follows:
\begin{align} \label{eq:cmab-mu-update}
\tilde{\mu}_{t,e}  =  X_{t,e}^{-1} b_{t,e}
\end{align}
To consider an upper confidence bound on the estimated utility of  an action at time $t$, we consider the following confidence bound  $\text{CB}_{t,e}$ \cite{auer2010ucb-revisited}:
\begin{align} \label{eq:cmab-confidence-bound}
\text{CB}_{t,e}(x_{t,a})  = \alpha \sqrt{x_{t,a}^\top X_{t,a}^{-1}x_{t,a} \log (t+1)}
\end{align}
where $\alpha$ is the learning parameter that controls the exploration for new actions. It should be noted that the $x_{t,a}^\top X_{t,a}^{-1}x_{t,e}$ term in (\ref{eq:cmab-confidence-bound}) accounts for the inverse of the number of times for which an action has been selected up until time $t$. In other words, the more the action is selected, the more confident is the estimate of its  utility, and thus the lower the inflation of its utility, i.e., $\text{CB}_{t,e}$. To select the best action at time epoch $t$, the learning agent should find the action that maximizes the following:
\begin{align} \label{eq:cmab-best-action}
a_t  = \underset{a \in \Gamma}{\arg\max} \qquad \bar{\mu}^\top_{t}x_{t,a} + \tilde{\mu}^\top_{t,e}x_{t,a}   + \text{CB}_{t,e}(x_{t,a})
\end{align}

\begin{figure*}
\centering
\includegraphics[scale=.4]{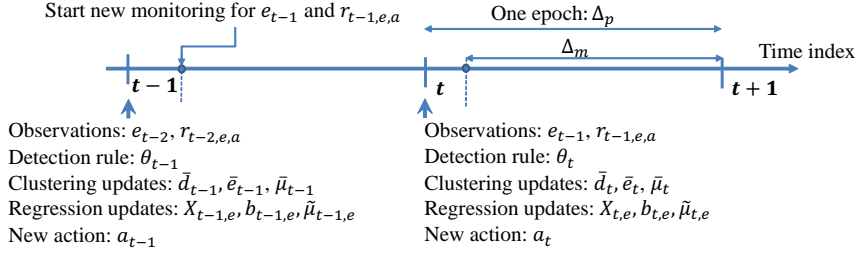}
\caption{Timing of different updates over a sequence of time epochs.} \label{fig:cmab-updates}
\end{figure*}

A pseudo-code for the algorithm is given in Algorithm \ref{alg:CMAB}. The algorithm starts by determining the best cluster that represents the environment. To do so, the device monitors the unlicensed channel for $\Delta_m$ period and constructs an initial SF $e_0$ that represents the initial state of the environment. The device finds the best cluster according to (\ref{eq:k-mean}), and initializes the centroid $\bar{e}_0$, radius $\bar{d}_0$, and average actions utility $\bar{\mu}_0$, accordingly. The regression quantities $b_{0,e}$ and $X_{0,e}$ are initialized to zero vector and identity matrix. Fractional utility estimate $\tilde{\mu}_{0,e}$ is initialized to zero and the observed reward $r_{0,e,a}$ to zero. 

At the start of each time epoch, say epoch $t$, the device performs the following. First, it checks whether changes in environment require switching to a new cluster by running the detection rule in (\ref{eq:environment_change_dedetection_rule}). If $\theta_t$ is set, device updates centroid, radius, and average utility to match these of the new cluster and re-initializes $X_{t,e} = I$ and $b_{t,e}= \mathbf{0}$. If no substantial change is detected, the device maintains cluster parameters used over the previous epoch. Second, the device updates $X_{t,e}$, $b_{t,e}$, and $\tilde{\mu}_{t,e}$ according to (\ref{eq:cmab-X-update}), (\ref{eq:cmab-b-update}), and (\ref{eq:cmab-mu-update}), respectively. Third, the device finds the best action to be used over epoch $t$ by running the optimization in (\ref{eq:cmab-best-action}). Once the new action is selected, the device operates normally while monitoring the environment to obtain a new state representation, i.e., $e_t$, and reward $r_{t,e,a}$. The timing of different updates is illustrated in Figure \ref{fig:cmab-updates}.

\begin{algorithm}
\caption{CMAB Algorithm}\label{alg:CMAB}
\begin{algorithmic}[1]
\State \textbf{Input:} Number of clusters $N_c$; Set of ST values (actions), $\Gamma$; Exploration parameter $\alpha>0$; Number of time epochs, $T$; Set of average action utilities, $M_c$; Set of centroids of clusters, $C$; Set of radii of clusters, $D$.

\State \textbf{Initialization:} Monitor the unlicensed channel for $\Delta_m$ duration and construct an initial SF $e_0$. Find the best initial cluster, $k_0 = k^*$, by running k-means algorithm with respect to $e_0$ according to (\ref{eq:k-mean}): 
 
 $\bar{e}_0 = \bar{e}_{k^*}$, $\bar{d}_0 = \bar{d}_{k^*}$,  $\bar{\mu}_0 = \bar{\mu}_{k^*}$,
 
 $b_{0,e}= \mathbf{0} \in \mathbb{R}^{N_a}$, $X_{0,e}= \mathbf{I} \in \mathbb{R}^{N_a \times N_a}$,
 
 $\tilde{\mu}_{0,e} = \mathbf{0}^{N_a \times 1}$, and $r_{0,e,a}=0$.

\For{$\qquad t = 1, 2, \cdots, T$}

\State Run detection rule (\ref{eq:environment_change_dedetection_rule}) and obtain $\theta_{t}$

 \State \textbf{IF} $\theta_t  = 1$:
 
\hspace{0.25cm}  Find new cluster $k_t = k^{*}$ as in (\ref{eq:k-mean})

 \hspace{0.25cm} $\bar{e}_t = \bar{e}_{k^*}$, $\bar{d}_t = \bar{d}_{k^*}$,  $\bar{\mu}_t = \bar{\mu}_{k^*}$,
 
 \hspace{0.25cm} $b_{t,e}= \mathbf{0} $, $X_{t,e}= \mathbf{I}$, $\tilde{\mu}_{t,e} = \mathbf{0}$

\textbf{ELSE:}

 \hspace{0.25cm} $\bar{e}_t = \bar{e}_{t-1}$, $\bar{d}_t = \bar{d}_{t-1}$,  $\bar{\mu}_t = \bar{\mu}_{
 t-1}$,
 
 \hspace{0.25cm} $b_{t,e} = b_{t-1,e} + r_{t-1,e,a}x_{t-1,a}$, 
 
  \hspace{0.3cm} $X_{t,e}= X_{t-1,e} + x_{t-1,a}x_{t-1,a}^\top$,
 
 \hspace{0.3cm} $\tilde{\mu}_{t,e} = X^{-1}_{t,e}b_{t,e}$.

\State Find the best action, i.e., ST value:

$a_t  = \underset{a \in \Gamma}{\arg\max} \:\: \bar{\mu}^\top_{t}x_{t,a} + \tilde{\mu}_{t,e}^\top x_{t,e}+  \text{CB}_{t,e}(x_{t,a})$.

\State Use ST value $a_t$ to access the unlicensed channel for the rest of epoch duration, monitor a new reward, $r_{t,e,a}$, as in (\ref{eq:observed-reward}), and observe a new state of environment, $e_{t}$.

\EndFor
\end{algorithmic}
\end{algorithm}

\section{Performance Evaluation and Discussion}
\label{sec:evaluation}
\subsection{Simulation Setup}
\begin{figure}
 \centering
 
\includegraphics[scale=0.5]{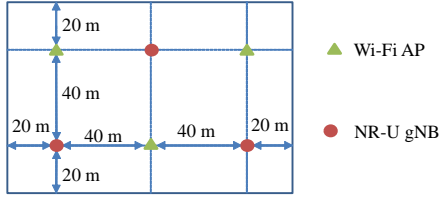}
 \caption{3GPP indoor simulation scenario ($a=20$ meter, $b=40$ meter)  \cite{TR38.889-nru-2}.} 
\label{fig:simulation-topology}
 \end{figure}

We develop a C++-based system-level simulator to study NR-U and Wi-Fi coexistence. Our simulator has been set with an accurate sub-nanosecond timing resolution. The simulator relies on a C++-library, called CSIM, that supports tracking of time, setting events, creating parallel processes, i.e., threads, as well as enabling communications between processes. For each device, we trigger parallel processes for handling various functions, including traffic generation, resource scheduling, LBT-based channel access, transmission over the air, etc. We implement the MAC layer for both NR-U and Wi-Fi networks and model the PHY layer as recommended by the 3GPP technical report \cite{TR38.889-nru-2} and IEEE 802.11 standard \cite{IEEE802.11-2016}. NR-U and Wi-Fi networks share a common channel of $20$ MHz bandwidth that is centered at 5.18 GHz. We consider the 3GPP indoor coexistence scenario, as shown in Figure \ref{fig:simulation-topology}. NR-U and Wi-Fi networks have 3 cells, and each cell serves 5 users. NR-U and Wi-Fi users are dropped in the simulation area uniformly while ensuring their minimum received power from their home cell is above $-82$ dBm. The channel is modeled according to 3GPP InH office channel model.

 Wi-Fi devices access the channel using the EDCA scheme \cite{IEEE802.11-2016}, and NR-U devices use the CAT4-LBT channel access scheme to access the shared channel \cite{TS37.213nru}. Unless stated otherwise, both networks serve FTP traffic with file size $0.5$ MB and Poisson arrival. We also analyze various traffic intensity for file arrivals. We also consider a random-walk mobility model for NR-U and Wi-Fi users with maximum speed of $1.5$ meters/sec. Unless stated otherwise, Wi-Fi devices access the channel according to the `Best Effort' access category (AC) with $W^{(w)}_{\min} = 16$, $W^{(w)}_{\max} = 1024$, and $a^{(w)} = 3$, while NR-U devices, on the other hand, access the channel according to the 3GPP `Priority Class 2' with $W^{(u)}_{\min} = 16$, $W^{(u)}_{\max} = 64$, and $a^{(u)} = 3$. The rest of simulation parameters, including shadowing, fading, traffic profile, etc., are set in line with settings in Annex A of 3GPP docs \cite{TR36.889LAA}\cite{TR38.889-nru-2}. 

To evaluate the gains of our framework, we compare Sense-Bandits framework with other two frameworks: Standard and Random framework. In the Standard framework, Wi-Fi and NR-U devices set their ST values according to their standard setting, i.e., $\gamma^{(w)}_j = -62$ dBm and   $\gamma^{(u)}_i = -72$ dBm. In the Random framework, we let NR-U and Wi-Fi devices select their ST values uniformly in the range $\{-82,-81, \cdots, -62\}$ dBm.

\subsection{Time Dynamics}

\begin{figure} 
  \centering
  \includegraphics[scale=.45]{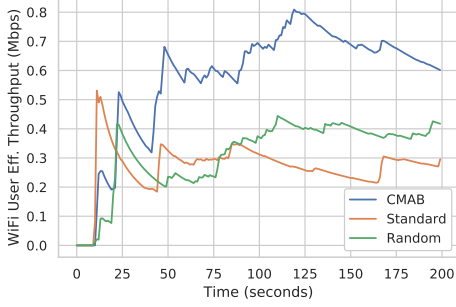}
  \caption{Wi-Fi user effective throughput vs. time.}    \label{fig:wifi_eff_throughput_vs_time}%
  \end{figure}

\begin{figure} 
  \centering
  \includegraphics[scale=.45]{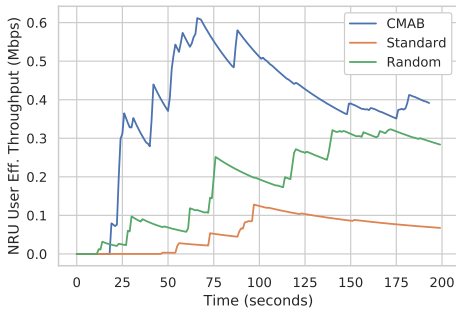}
  \caption{NR-U user effective throughput vs. time.}  
  \label{fig:nru_eff_throughput_vs_time}%
  \end{figure}

\textcolor{black}{We investigate the dynamics of CMAB algorithm over time when compared to other schemes, as shown in Figures \ref{fig:wifi_eff_throughput_vs_time} and \ref{fig:nru_eff_throughput_vs_time}. We evaluate the effective throughput achieved by the three schemes, i.e., CMAB, Standard, and Random. The effective throughout at a certain moment is computed by dividing the accumulated traffic delivered successfully by the duration of time up until the moment of interest. In Figures \ref{fig:wifi_eff_throughput_vs_time} and \ref{fig:nru_eff_throughput_vs_time}, we notice that the CMAB algorithm provides higher throughput than the other two schemes; thanks to the learning feature of CMAB for offering awareness about interference and dynamics of neighboring devices.}

\subsection{Performance Measures}
\textcolor{black}{We evaluate the throughput of CMAB algorithm against the Standard and Random schemes. In our evaluation, we divide users in each cell to two sets, users adapting their ST values, a.k.a., Adapting Nodes, and users fixing their ST values to the standard settings, a.k.a., Standard Nodes. In particular, we let the first three users in every cell adapt their ST values, while the remaining two users fix their ST values to the standard settings. This heterogeneous setup allows us to investigate the impact of adapting ST values under heterogeneous setup in which standard devices could exist in the proximity of the intelligent devices. We also report the throughput for both Adapting Nodes and Standard Nodes when the Adapting Nodes run other schemes as well. We investigate the User Perceived Throughput achieved by the three schemes at the MAC layer level. Throughput is calculated by dividing the file size by the time it takes to deliver the file to the receiver. Every file of 0.5 MB is divided into smaller segments (packet) with each segment consisting of 8 KB.} 
\subsubsection{NR-U Performance}

\begin{figure} 
  \centering
  \includegraphics[scale=.4]{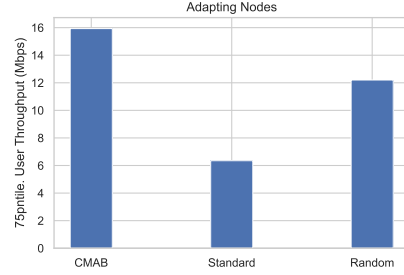}
  \caption{75\% percentile of NR-U user perceived throughput for users adapting their ST values.}    \label{fig:nru_upt_75p}%
  \end{figure}

\begin{figure} 
  \centering
  \includegraphics[scale=.4]{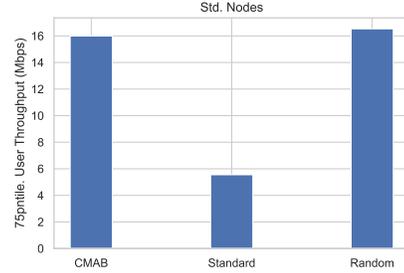}
  \caption{75\% percentile of NR-U user perceived throughput for users using fixed standard-compliant ST values.}    \label{fig:nru_std_upt_75p}%
  \end{figure}

In Figures \ref{fig:nru_upt_75p} and \ref{fig:nru_std_upt_75p}, we plot the 75th percentile of throughput for NR-U Adapting Nodes and Standard Nodes, respectively, achieved under the three schemes (i.e., CMAB, Standard, and Random). We observe that CMAB algorithm provides NR-U Adapting Nodes higher throughput than the Standard and Random schemes. We also observe an exciting observation regarding the Random scheme. It can be observed that random assignment of ST values can still provide higher throughput when compared to the standard settings. We also observe that the Standard Nodes experience better throughput when they share the channel with the Adapting Nodes under the CMAB scheme when compared to all-standard setting in which all devices stick to their standard ST values. This proves that adapting ST values improves the performance of both Adapting Nodes and standard-complaint ones as well.

\subsubsection{Wi-Fi Performance}

\begin{figure} 
  \centering
  \includegraphics[scale=.4]{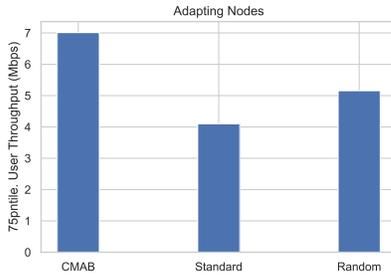}
  \caption{75\% percentile of Wi-Fi user perceived throughput for users adapting their ST values.}    \label{fig:wifi_upt_75p}%
  \end{figure}

\begin{figure} 
  \centering
  \includegraphics[scale=.4]{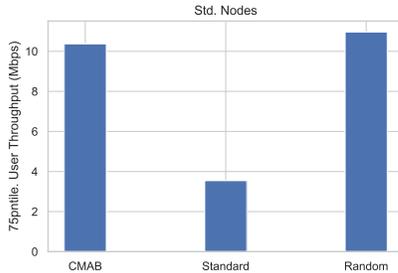}
  \caption{75\% percentile of Wi-Fi user perceived throughput for users using fixed standard-compliant ST values.}    \label{fig:wifi_std_upt_75p}%
  \end{figure}

In Figures \ref{fig:wifi_upt_75p} and \ref{fig:wifi_std_upt_75p}, we plot the 75th percentile of Wi-Fi throughput for Adapting Nodes and Standard Nodes, respectively, under the three schemes. We observe the CMAB algorithm provides Adapting Nodes higher throughput than the Standard and Random schemes. Once again, it can be observed that the random assignment of ST values still provides higher throughput than the standard settings. The Wi-Fi Standard Nodes experience better throughput when they share the channel with the Adapting Nodes under the CMAB scheme as well.

\section{Conclusions} \label{sc:conclusions}

In this work, we introduced a new framework for distributed learning and adaptation of ST values used by NR-U and Wi-Fi devices. The new framework relies on a novel clustering-based multi-armed bandit (CMAB) algorithm. We conducted extensive system-level simulations to evaluate the gains and performance improvement of CMAB as compared to fixed standard as well as random setting of ST values. We spotted many exciting observations. The CMAB algorithm provided higher throughput than the standard settings. In addition, we found that even the random adaptation of ST values could provide higher throughput than fixed settings. The CMAB algorithm acted as a friendly neighbor to standard-compliant devices that do not adapt their ST values.




\bibliographystyle{IEEEtran}
\bibliography{icccn-2021}

\begin{thebibliography}{10}
\providecommand{\url}[1]{#1}
\csname url@samestyle\endcsname
\providecommand{\newblock}{\relax}
\providecommand{\bibinfo}[2]{#2}
\providecommand{\BIBentrySTDinterwordspacing}{\spaceskip=0pt\relax}
\providecommand{\BIBentryALTinterwordstretchfactor}{4}
\providecommand{\BIBentryALTinterwordspacing}{\spaceskip=\fontdimen2\font plus
\BIBentryALTinterwordstretchfactor\fontdimen3\font minus
  \fontdimen4\font\relax}
\providecommand{\BIBforeignlanguage}[2]{{%
\expandafter\ifx\csname l@#1\endcsname\relax
\typeout{** WARNING: IEEEtran.bst: No hyphenation pattern has been}%
\typeout{** loaded for the language `#1'. Using the pattern for}%
\typeout{** the default language instead.}%
\else
\language=\csname l@#1\endcsname
\fi
#2}}
\providecommand{\BIBdecl}{\relax}
\BIBdecl

\bibitem{ITU-IMT2020-Chairman-Notes}
ITU, ``{Chairmans report},'' no. FG IMT-2020, Dec. 2016.

\bibitem{TR36.889LAA}
3GPP, ``Study on licensed-assisted access to unlicensed spectrum,'' 3GPP TR.
  36.889 v13.0.0., Jun. 2015.

\bibitem{TR38.889-nru-2}
------, ``{Study on NR--based access to unlicensed spectrum},'' no. 3GPP TR
  38.889 v16.0.0, Dec. 2018.

\bibitem{Lagen2018nru-mmwave}
S.~Lagen, L.~Giupponi, S.~Goyal, N.~Patriciello, B.~Bojovic, A.~Demir,
  M.~Beluri, and J.~Mangues-Bafalluy, ``New radio beam-based access to
  unlicensed spectrum: Design challenges and solutions,'' \emph{arXiv preprint
  arXiv:1809.10443}, 2018.

\bibitem{Hirzallah2020TCCN-nru}
M.~{Hirzallah}, M.~{Krunz}, B.~{Kecicioglu}, and B.~{Hamzeh}, ``{5G} new radio
  unlicensed: Challenges and evaluation,'' \emph{IEEE Transactions on Cognitive
  Communications and Networking}, pp. 1--1, 2020.

\bibitem{Jiang2007tmc-hidden-exposed-node}
L.~B. {Jiang} and S.~C. {Liew}, ``Improving throughput and fairness by reducing
  exposed and hidden nodes in 802.11 networks,'' \emph{IEEE Transactions on
  Mobile Computing}, vol.~7, no.~1, pp. 34--49, Jan 2008.

\bibitem{Hirzallah2017jsac}
M.~Hirzallah, W.~Afifi, and M.~Krunz, ``Full-duplex-based rate/mode adaptation
  strategies for {Wi-Fi/LTE-U} coexistence: A {POMDP} approach,'' \emph{IEEE
  Journal on Selected Areas in Communications}, vol.~35, no.~1, pp. 20--29, Jan
  2017.

\bibitem{Amir2019Dyspan}
A.~H.~Y. Abyaneh, M.~Hirzallah, and M.~Krunz, ``Intelligent-{CW}: {AI}-based
  framework for controlling contention window in {WLANs},'' in \emph{Proc. of
  IEEE International Symposium on Dynamic Spectrum Access Networks (DySPAN)},
  2019, pp. 1--10.

\bibitem{Hirzallah2019Secon-matchmaker}
M.~{Hirzallah}, Y.~{Xiao}, and M.~{Krunz}, ``Matchmaker: An inter-operator
  network sharing framework in unlicensed bands,'' in \emph{Proc. of IEEE
  SECON'19 Conference}, June 2019, pp. 1--9.

\bibitem{Hirzallah2020icnc}
M.~Hirzallah and M.~Krunz, ``Intelligent tracking of network dynamics for
  cross-technology coexistence over unlicensed bands,'' in \emph{Proc. of
  International Conference on Computing, Networking and Communications (ICNC)},
  2020, pp. 698--703.

\bibitem{Li2017tvt-laa-ed-adaptation}
L.~{Li}, J.~P. {Seymour}, L.~J. {Cimini}, and C.~{Shen}, ``Coexistence of
  {Wi-Fi} and {LAA} networks with adaptive energy detection,'' \emph{IEEE
  Transactions on Vehicular Technology}, vol.~66, no.~11, pp. 10\,384--10\,393,
  Nov 2017.

\bibitem{Iqbal2017laa-ed-ns3}
M.~{Iqbal}, C.~{Rochman}, V.~{Sathya}, and M.~{Ghosh}, ``Impact of changing
  energy detection thresholds on fair coexistence of {Wi-Fi} and {LTE} in the
  unlicensed spectrum,'' in \emph{Proc. of Wireless Telecommunications
  Symposium (WTS)}, April 2017, pp. 1--9.

\bibitem{Ajami2017twc-11ax-stochastic-geometry}
A.~{Ajami} and H.~{Artail}, ``On the modeling and analysis of uplink and
  downlink {IEEE} 802.11ax {Wi-Fi} with {LTE} in unlicensed spectrum,''
  \emph{IEEE Transactions on Wireless Communications}, vol.~16, no.~9, pp.
  5779--5795, Sep. 2017.

\bibitem{Mehrnoush2018tnet-laa-modeling}
M.~{Mehrnoush}, V.~{Sathya}, S.~{Roy}, and M.~{Ghosh}, ``Analytical modeling of
  {Wi-Fi} and {LTE-LAA} coexistence: Throughput and impact of energy detection
  threshold,'' \emph{IEEE/ACM Transactions on Networking}, vol.~26, no.~4, pp.
  1990--2003, Aug 2018.

\bibitem{kulkarni2016dynamic}
P.~Kulkarni and F.~Cao, ``Dynamic sensitivity control to improve spatial reuse
  in dense wireless {LANs},'' in \emph{Proc. of the ACM MSWiM Conference},
  2016, pp. 323--329.

\bibitem{Mvulla2018-access-dual-sensing}
J.~{Mvulla} and E.~{Park}, ``Enhanced dual carrier sensing with transmission
  time control for fair spatial reuse in heterogeneous and dense {WLANs},''
  \emph{IEEE Access}, vol.~6, pp. 22\,140--22\,155, May 2018.

\bibitem{Oteri2015globecom-spatial-reuse}
O.~{Oteri}, F.~{La Sita}, R.~{Yang}, M.~{Ghosh}, and R.~{Olesen}, ``Improved
  spatial reuse for dense 802.11 {WLANs},'' in \emph{Proc. of IEEE Globecom
  Workshops (GC Wkshps)}, Dec 2015, pp. 1--6.

\bibitem{afaqui2016globecom-cca-adaptation}
M.~S. {Afaqui}, E.~{Garcia-Villegas}, E.~{Lopez-Aguilera}, and D.~{Camps-Mur},
  ``Dynamic sensitivity control of access points for {IEEE} 802.11ax,'' in
  \emph{Proc. of IEEE ICC Conference}, May 2016, pp. 1--7.

\bibitem{selinis2019damysus}
I.~Selinis, K.~Katsaros, S.~Vahid, and R.~Tafazolli, ``Damysus: A practical
  {IEEE} 802.11 ax {BSS} color aware rate control algorithm,''
  \emph{International Journal of Wireless Information Networks}, vol.~26,
  no.~4, pp. 285--307, 2019.

\bibitem{wilhelmi2019-ml-cca}
F.~Wilhelmi, S.~Barrachina-Mu{\~n}oz, B.~Bellalta, C.~Cano, A.~Jonsson, and
  G.~Neu, ``Potential and pitfalls of multi-armed bandits for decentralized
  spatial reuse in wlans,'' \emph{Journal of Network and Computer
  Applications}, vol. 127, pp. 26--42, 2019.

\bibitem{wilhelmi2019collaborative}
F.~Wilhelmi, C.~Cano, G.~Neu, B.~Bellalta, A.~Jonsson, and
  S.~Barrachina-Mu{\~n}oz, ``Collaborative spatial reuse in wireless networks
  via selfish multi-armed bandits,'' \emph{Ad Hoc Networks}, vol.~88, pp.
  129--141, 2019.

\bibitem{TS37.213nru}
3GPP, ``Physical layer procedures for shared spectrum channel access,'' 3{GPP}
  {TS}. 37.213 v16.5.0., Mar. 2021.

\bibitem{IEEE802.11-2016}
{IEEE}, ``{IEEE}--part 11: Wireless {LAN MAC} and {PHY} layer specifications,''
  pp. 1--3534, 2016.

\bibitem{Hirzallah2019tccn}
M.~{Hirzallah}, M.~{Krunz}, and Y.~{Xiao}, ``Harmonious cross-technology
  coexistence with heterogeneous traffic in unlicensed bands: Analysis and
  approximations,'' \emph{IEEE Transactions on Cognitive Communications and
  Networking}, vol.~5, no.~3, pp. 690--701, Sep. 2019.

\bibitem{abbasi2011improved-MAB-bounds}
Y.~Abbasi-Yadkori, D.~P{\'a}l, and C.~Szepesv{\'a}ri, ``Improved algorithms for
  linear stochastic bandits,'' in \emph{Advances in Neural Information
  Processing Systems}, 2011, pp. 2312--2320.

\bibitem{auer2010ucb-revisited}
P.~Auer and R.~Ortner, ``{UCB} revisited: Improved regret bounds for the
  stochastic multi-armed bandit problem,'' \emph{Periodica Mathematica
  Hungarica}, vol.~61, no. 1-2, pp. 55--65, 2010.

\end{thebibliography}
%




\end{document}